\documentclass[oldversion]{aa} 
\usepackage{graphicx}
\usepackage{txfonts}
\usepackage[]{natbib}
\usepackage{aalongtable}
\usepackage{lscape}
\usepackage[figuresright]{rotating}
\usepackage{color}
\begin{document}
   \title{Radio observations of the \emph{Chandra} Deep Field South}

   \subtitle{Exploring the possible link between radio emission and star formation in X-ray selected AGN}

   \authorrunning{Rovilos et al.}

   \author{E. Rovilos,\inst{1,2} 
           A. Georgakakis,\inst{3}
           I. Georgantopoulos,\inst{1}
           J. Afonso,\inst{4}
           A. M. Koekemoer,\inst{5}
           B. Mobasher,\inst{5}
           \and
           C. Goudis\inst{1,2}
          }

   \offprints{E. Rovilos\\ \email{erovilos@astro.noa.gr}}

   \institute{Institute for Astronomy and Astrophysics, National
             Observatory of Athens, I. Metaxa \& V. Pavlou str, Palaia
             Penteli, 15236, Athens, Greece
         \and
             Astronomical Laboratory, Department of Physics, University
             of Patras, 26500, Rio-Patras, Greece
         \and
             Astrophysics Group, Imperial College, Prince Consort Rd.,
             SW7 2AZ, London, United Kingdom
         \and
             Centro de Astronomia da Universidade de Lisboa, Observat\'{o}rio
             Astron\'{o}mico de Lisboa, 1349-018 Lisboa, Portugal
         \and
             Space Telescope Science Institute, 3700 San Martin Drive,
             Baltimore, MD 21218, USA
             }

   \date{Received date; accepted date}

\abstract{ We explore the nature of the radio emission of X-ray
selected AGN by combining deep radio (1.4\,GHz; 60\,$\mu$Jy) and X-ray
data with multiwavelength (optical, mid-infrared) observations in the
Extended Chandra Deep Field South (E-CDFS). The fraction of radio
detected X-ray sources increases from 9\% in the E-CDFS to 14\% in the
central region of this field, which has deeper X-ray coverage from the
1\,Ms CDFS. We find evidence that the radio emission of up to 60\% of
the hard X-ray/radio matched AGN is likely associated with
star-formation in the host galaxy. Firstly, the mid-IR (24\,$\mu$m)
properties of these sources are consistent with the infrared/radio
correlation of starbursts. Secondly, most of them are found in
galaxies with blue rest-frame optical colours ($U-V$), suggesting a
young stellar population. On the contrary, X-ray/radio matched AGN
which are not detected in the mid-infrared have red $U-V$ colours
suggesting their radio emission is associated with AGN activity. We
also find no evidence for a population of heavily obscured
radio-selected AGN that are not detected in X-rays. Finally, we do no
confirm previous claims for a correlation between radio emission and
X-ray obscuration. Assuming that the radio continuum measures
star-formation, this finding is against models where the dust and gas
clouds associated with circumnuclear starbursts are spherically
blocking our view to the central engine. 
}


    \maketitle
%

\section{Introduction}

A major recent development in extragalactic astronomy is the discovery
that a large fraction of the spheroids in the local Universe contain a
super-massive black hole \citep[e.g.][]{Magorrian1998} and that the
mass of this monster is tightly correlated to the stellar mass of host
galaxy bulge \citep[e.g.][]{Ferrarese2000,Gebhardt2000}. The
implication of this fundamental observation is that the formation and
evolution of galaxies and the build-up of super-massive BHs at their
centers are interconnected. The evidence above has prompted attempts
to model the interplay between black-hole growth and star-formation
\citep[e.g.][]{Fabian1998}. A recent development in this direction are
simulations which include AGN feedback mechanisms
\citep[e.g.][]{Hopkins2005}. In these, galaxy mergers trigger the AGN
and also produce nuclear starbursts that both feed and obscure the
central engine for most of its active lifetime
($\approx 10^{8}$\,yr). AGN-driven outflows also develop,
which at later stages become strong enough to rapidly quench the
star-formation, leaving behind a red passive remnant and allowing the
AGN to shine unobscured for a short period
($\approx 10^{7}$\,yr). The attractiveness of this class
of models is that they reproduce a number of AGN properties (e.g.
duty-cycle, $N_{\mathrm{H}}$ distribution, luminosity function), while
they can potentially explain galaxy properties, such as the bimodality
of the colour magnitude diagram
\citep[e.g.][]{Bell2004,Willmer2006,Nandra2006}, via the AGN-driven
regulation of the star-formation.

The predictions of the AGN feedback model above include a link between
the obscured AGN stage and starburst events. In the local Universe
$(z<0.1)$ an association between {\it optically} selected AGN and
recent/ongoing star-formation has been established using the Sloan
Digital Sky Survey \citep[SDSS;][]{Kauffmann2003}. The link between
obscured accreting BHs and star-formation remains controversial
however, despite claims that Seyfert-IIs show optical spectroscopic
evidence for young stellar populations
\citep[e.g.][]{CidFernandes2001}. The low-$z$ results above, although
broadly consistent with the merger formation models of
\citet{Hopkins2005}, are based on optically selected AGN samples,
which may miss a substantial fraction of the BH accretion in the
Universe \citep[e.g.][]{Mushotzky2004}. At higher redshift in
particular ($z\approx 1$), close to peak of the global AGN density
\citep*[e.g.][]{Barger2005,Hasinger2005}, X-ray observations are
arguably the most efficient tool for identifying AGN over a wide range
of luminosities and obscurations. The stellar population of the X-ray
selected AGN in deep surveys is still under
debate. \citet{Franceschini2005} find that about 40\% of the X-ray
sources in the ELAIS-N1 SWIRE region have mid-IR SEDs consistent with
ongoing star-formation. Contamination of these wavelengths from hot
dust emission related to the central accreting BH remains an issue
however, when interpreting the mid-IR properties of X-ray
sources. Recently, \citet{Nandra2006} find that the majority of the
$z\approx 1$ X-ray sources in the AEGIS survey \citep{Davis2006} have
rest-frame colours of passive red galaxies indicating a dominant old
stellar population with little, if any, current star-formation.

One of the difficulties in studying the stellar content of AGN hosts
is the decomposition of the stellar light from the emission of the
accreting BH. Combining information from different parts of the
electromagnetic spectrum is essential to address this issue. In this
paper we follow such a multiwavelength approach to study the nature of
the $\mu$Jy radio emission of X-ray sources in the Extended Chandra
Deep Field South (E-CDFS) and to explore the implications in the
context of the AGN/star-formation connection. This is motivated by
previous claims that X-ray selected type-II (X-ray absorbed) AGN
are more frequently associated with $\mu$Jy flux density radio sources
\citep{Bauer2002,Georgakakis2004}. At these faint limits the radio
population is dominated by starbursts
\citep{Georgakakis1999,Chapman2003,Afonso2005} and therefore,
the above finding has been tentatively interpreted as evidence for a
link between star-formation and obscured AGN. The radio luminosities
of these systems ($\approx 10^{23}$\,W\,Hz$^{-1}$) are well below
those of the classic radio galaxies ($>10^{24}$\,W\,Hz$^{-1}$),
suggesting that star-formation may contribute or even dominate the
observed radio emission.
 
In addition to the deep X-ray observations in the E-CDFS the
multiwavelnegth data used here include ultra-deep wide-area radio
1.4\,GHz observations (60\,$\mu$Jy; 0.3\,deg$^{2}$;
\citealt{Afonso2006}), Spitzer mid-IR photometry as part of the Great
Observatories Origins Deep Survey (GOODS) as well as extensive optical
photometric and spectroscopic follow-up programs. These complementary
multiwavelength observations are essential to explore the nature of
the $\mu$Jy radio emission in X-ray selected AGN. Throughout this
paper we adopt $\mathrm{H_{0}=72\,km\,s^{-1}\,Mpc^{-1}}$,
$\Omega_{\mathrm{M}}=0.3$ and $\Omega_{\Lambda}=0.7$.


\section{The Data}

  \begin{figure}
  \centering
  \includegraphics[width=8.8cm]{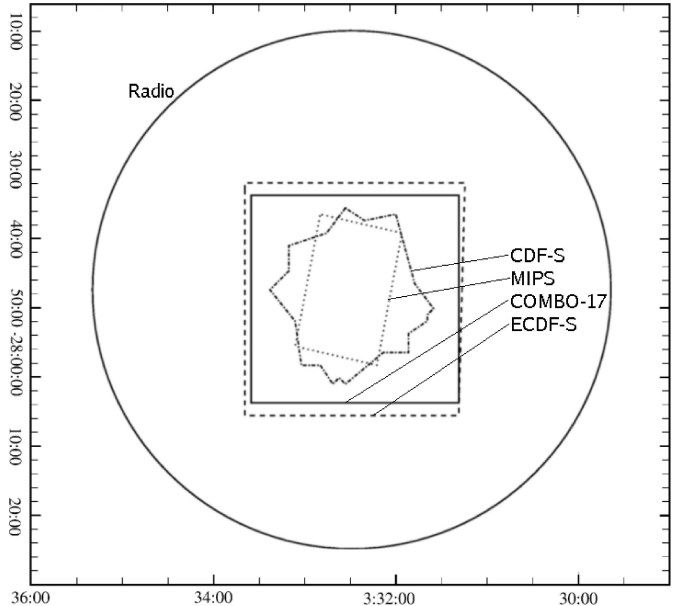}
  \caption{Layout of the
           multiwavlength observations in the
           E-CDFS. The dot-dashed and dashed lines delineate the 1\,Ms CDFS
           and the shallower 250\,ks E-CDFS respectively. The circle marks
           the limits of the ATCA radio observations. The continuous-lined
           square marks the boundaries of the COMBO-17 survey and the
           the dotted line the area covered by the Spitzer/MIPS
           observations.}
  \label{fields}
  \end{figure}

The X-ray observations are from the 1\,Ms CDFS and the E-CDFS. These
two surveys cover a total contiguous area of 0.3\,deg$^2$ to a
limiting $2-8$\,keV flux of $\approx 2.8\times
10^{-16}$\,erg\,cm$^{-2}$\,s$^{-1}$ at the most sensitive central
region of the CDFS increasing to $\approx 6.7\times
10^{-16}$\,erg\,cm$^{-2}$\,s$^{-1}$ in the E-CDFS. In this paper we
use the X-ray point source catalogs constructed by
\citet{Alexander2003} and \citet{Lehmer2005} for the CDFS and E-CDFS
respectively.

The radio data are presented in \citet{Afonso2006} and more analytically
in Koekemoer et al. (in preparation). In brief, the observations are
performed at 1.4\,GHz using the Australia Telescope Compact Array (ATCA)
and cover a total area of 1.2\,deg$^2$ that includes both the CDFS and the
E-CDFS (see Fig. \ref{fields}).
120 hours of data were obtained, causing
the $1\sigma$ rms noise level to range from 14\,$\mu$Jy at the central part
to 100\,$\mu$Jy at the outer region, while the beam size used for imaging is
$(16.8\times6.95)$\,arcsec. Its shape is a result of the ATCA configuration,
which is more spread in the E-W direction, and this reflects also to the
positional uncertainties, which are larger in declination. The sources
are selected using the False Discovery Rate method. The final list
comprises 681 sources with integrated flux densities ranging from
61\,$\mu$Jy to 170\,mJy. A total of 133 and 337 sources overlap with the
CDFS and the E-CDFS respectively.

Mid-infrared MIPS-24\,$\mu$m data are available as part of the Great
Observatories Origins Deep Survey (GOODS) Legacy Survey. The
observations were carried out by the Spitzer Space Telescope and cover
a total area of about $10\times 16.5\,\mathrm{arcmin}^2$ centered on
the CDFS (see Fig. \ref{fields}). We have used the public
catalog\footnote{\texttt{http://data.spitzer.caltech.edu/popular/goods/}},
which contains 948 sources to the limit of 80\,$\mu$Jy.

Multi-waveband optical/near-IR photometric as well as optical
spectroscopic observations have been carried out by many groups in the
CDFS and E-CDFS survey regions, providing redshift estimates
(spectroscopic or photometric) for different source populations. The
spectroscopic redshifts used here are from \citet{Szokoly2004},
\citet{Vanzella2005,Vanzella2006}, \citet{LeFevre2004} and
\citet{Mignoli2005}. For sources without spectroscopic redshifts we
rely on photometric redshifts mainly from \citet{Wolf2004} as part of
the COMBO-17 survey. This uses 17 photometric passband and three sets
of template SED (stars, galaxies and QSOs) to determine the redshifts
of $\sim 65000$ objects with an accuracy of a few percent (up to 10\%
for $R\gtrsim 24$). The area covered by COMBO-17 covers the CDFS and
most of the E-CDFS (see Fig. \ref{fields}). Photometric redshifts
from \citet{Zheng2004} are also used in the case of CDFS X-ray sources
fainter than the COMBO-17 magnitude limit.

\section{The Sample}

Given the positional uncertainties of the X-ray and radio sources we
search for counterparts in a radius corresponding to the $3\sigma$
radio positional uncertainty. For most cases the radio/X-ray offset is
less than 3\,arcsec. We
find 44 and 72 matches for the CDFS and the E-CDFS respectively with
27 common identifications. We quantify the fraction of spurious
matches using mock catalogs produced by randomly shifting the
positions of X-ray and radio sources adopting a maximum offset of
0.5\,arcmin. The fraction of spurious identifications is estimated to
be 1.8\% and 2.4\% of the X-ray/radio matched population for the CDFS
and E-CDFS respectively. Searching for mid-infrared counterparts for
CDFS sources using a 3\,arcsec search radius returned 133 matches, 32
of which have radio counterparts. Searching for matches among the
radio sources which are inside the CDFS (133 sources) returned 52
matches.

The radio detection rates for the CDFS and E-CDFS X-ray sources are
thus 14\% and 9\% respectively, while the X-ray detection rates for
radio sources are 33\% for the CDFS and 21\% for the E-CDFS case. The
radio and X-ray detection rates measured in \citet{Bauer2002} are 26\%
and 71\% respectively for the 1\,Ms \emph{Chandra} Deep Field North
(CDF-N). The 1\,Ms CDF-N X-ray catalog has a flux limit of $\approx
1\times10^{-16}$\,erg\,cm$^{-2}$\,s$^{-1}$, marginally fainter than
the CDFS ($1.3\times 10^{-16}$\,erg\,cm$^{-2}$\,s$^{-1}$ in the full
band). The radio flux density limit in \citet{Bauer2002} is $\approx
40\,\mu$Jy, while our detection threshold in the central region is
$\approx 60\,\mu$Jy. It is interesting to note that both the X-ray and
radio detection rates increase with deeper
observations. \citet{Bauer2002} report that a large fraction of their
radio-X-ray matched sources lie close to their detection limit, so
deeper radio observations would increase the detection rate.

Table \ref{table} lists the basic X-ray, radio and infrared
properties of the 89 unique matches. the columns are:
\begin{enumerate}
\item{Identification number in the \citet{Alexander2003} catalog.}
\item{Identification number in the \citet{Lehmer2005} catalog.}
\item{X-ray position Right Ascension (J2000).}
\item{X-ray position Declination (J2000).}
\item{Spectroscopic (if available) or photometric redshift.}
\item{Hard band ($2.0-8.0$\,keV) X-ray flux.
      The fluxes have been calculated from the hard-band
      count rates assuming a power-law spectrum with
      an intrinsic photon index $\Gamma=1.9$
      \citep{Nandra1994}
      and are corrected for absorption.}
\item{Hard band X-ray rest-frame luminosity based on the above
      fluxes and corrected for intrinsic absorption.}
\item{Hardness ratio, $HR=\frac{H-S}{H+S}$, where $H$ and $S$ are the count
      rates in the hard ($2-8$\,keV) and soft ($0.5-2$\,keV) bands
      respectively.}
\item{Hydrogen column density calculated using the hardness ratio of
      each source and assuming an intrinsic power-law X-ray spectrum with
      $\Gamma=1.9$ and a Galactic column density of
      $N_{\rm H}=10^{20}$\,cm$^{-2}$ appropriate for the CDFS.}
\item{Radio (rest-frame) luminosity, assuming a radio spectral index of
      0.8 $(f_{\rm 1.4\,GHz}\propto\nu^{-\alpha})$.
      This may overestimate the radio luminosity of AGN dominated systems,
      as their radio spectrum is flatter because of synchrotron self
      absorption \citep[e.g.][]{Taylor1996}}
\item{24\,$\mu$m rest frame luminosity. We have used the SED
      of NGC\,1068 \citep{Lutz1997}, a typical Seyfert-2, to perform the
      K-correction. Using starburst galaxy SEDs, such as M\,82
      or Arp\,220, would result in larger K-corrections at $z\approx1.4$,
      because of the strong absorption from PAHs and
      silicate features at $\lambda\approx 10\,\mu$m.}
\item{Source classification based on the X-ray luminosity (not
      corrected for absorption) and the hardness ratio according to
      scheme:
      \begin{itemize}
      \item{Galaxies, if $L_{\rm x}<10^{42}$\,erg\,s$^{-1}$ and
            $HR\leq-0.2$,}
      \item{AGN2, if $HR>-0.2$,}
      \item{AGN1, if $L_{\rm x}\geq 10^{42}$\,erg\,s$^{-1}$ and
            $HR\leq-0.2$.}
      \end{itemize}
      These criteria are similar to those adopted by
      \citet{Bauer2004}. The classification of individual
      sources presented here is 
      consistent with that study. Sources in the normal galaxy group will
      not be discussed in the rest of the paper. 
      }
\end{enumerate}

\section{Results}

\subsection{The stellar population of AGN}\label{radio-ir}

  \begin{figure}
  \centering
  \includegraphics[width=8.8cm]{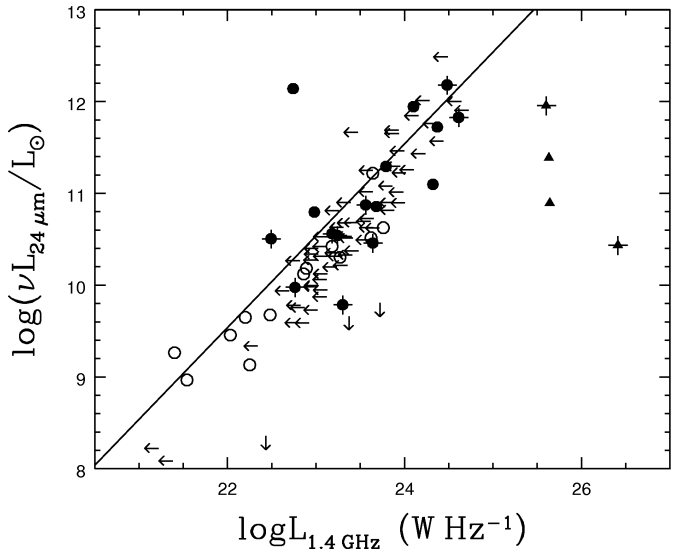}
  \caption{24\,$\mu$m versus radio luminosity. AGN are shown as filled
           circles and radio loud AGN (with
           $\log L_{\rm 1.4\,GHz}>10^{25}$\,W\,Hz$^{-1}$) as filled triangles.
           Crosses mark obscured objects (AGN2). The line is the best fit
           $L_{\rm 24\,\mu m}-L_{\rm 1.4\,GHz}$ for Spitzer starburst
           galaxies \citep{Wu2005}. Open circles represent radio/infrared
           sources without an X-ray counterpart.}
  \label{l24lr}
  \end{figure}

  \begin{figure}
  \centering
  \includegraphics[width=8.8cm]{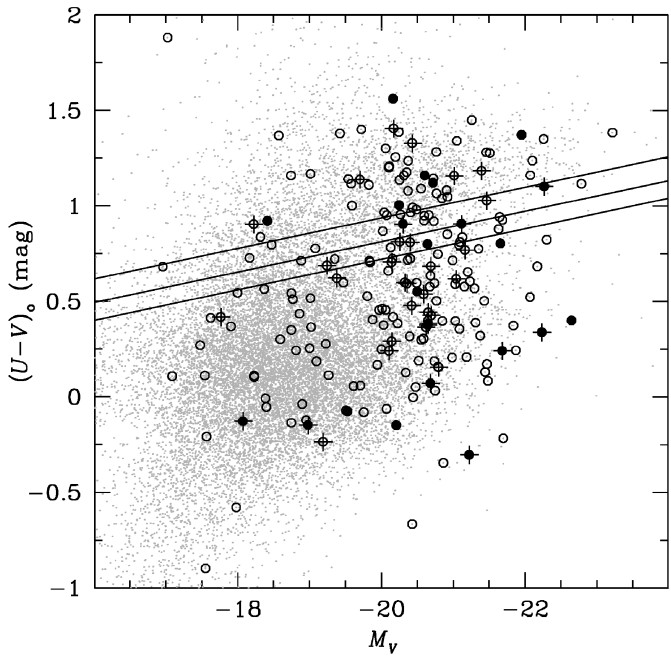}
  \caption{Colour-magnitude diagram for hard X-ray selected AGN
           in the redshift interval $0.2<z<1.2$ probed by the
           COMBO-17 survey. Open circles are X-ray sources without radio
           counterparts while filled circles correspond to
           X-ray/radio matches. X-ray sources detected at 24\,$\mu$m
           are marked with a cross. 
           Optically selected galaxies in the same redshift range
           from the COMBO-17 photometric survey are plotted as gray dots. 
           The lines mark the borders of the red cloud for redshifts 0.5,
           0.9 and 1.2, according to the relation
           $\langle U-V\rangle=1.15-0.31z-0.08(M_{\mathrm{V}}-5\log h+20)$
           with $h=0.72$ \citep{Bell2004}.}
  \label{cmd}
  \end{figure}

Figure \ref{l24lr} plots the 24\,$\mu$m MIR luminosity against
1.4\,GHz radio power. For starburst galaxies these two quantities are
correlated \citep{Appleton2004,Wu2005}, as both wavelengths are
believed to probe the same population of young massive stars. In this
figure, many X-ray/radio matched AGN (up to 60\% of the hard
X-ray/radio population within the MIPS area) scatter around the
MIR/1.4\,GHz relation of star-forming galaxies \citep{Wu2005},
suggesting ongoing star-formation is likely to dominate the observed
radio emission. Moreover, X-ray sources not detected to the limit of
the radio survey have $L_{\rm 1.4\,GHz}$ upper limits which are also
broadly consistent with that relation. This is tentative evidence that
a substantial fraction of the overall X-ray selected AGN population
have MIR and radio properties consistent with star-formation in the
host galaxy. The radio observations are not deep enough to detect
these systems, although Spitzer/MIPS is sufficiently sensitive to
identify this population at 24\,$\mu$m. In Fig. \ref{l24lr} there
are also X-ray/radio matched AGN that deviate from the MIR/1.4\,GHz
relation of star-forming galaxies. These include sources without
24\,$\mu$m counterparts, plotted as upper limits and radio loud AGN,
defined as sources with $\log L_{\rm 1.4\,GHz}>10^{25}$\,W\,Hz$^{-1}$
\citep{Kellermann1989}. The radio emission of both these populations
is associated with accretion on a supper-massive BH.

\citet{Donley2005} report a population of radio selected AGN not
detected in X-rays but identified by their excess radio to infrared emission.
Figure \ref{l24lr} plots the infrared/radio sources not identified
with the 1\,Ms exposure of the CDFS with open circles. There is no
significant deviation of those sources from the correlation line,
suggesting they, too, are of starburst nature. We therefore cannot
confirm the result of \citet{Donley2005} based on radio sources
detected by MIPS. There are however radio sources without neither an
X-ray nor a MIPS detection which can be considered being of radio
excess based on their infrared upper limits. See also recent results
from \citet*{Barger2007}.

A complementary approach for studying the stellar population of AGN
hosts is to use the colour-magnitude diagram (CMD), which provides a
powerful diagnostic for separating evolved from star-forming galaxies
\citep[e.g.][]{Bell2004,Willmer2006,Nandra2006}. Figure \ref{cmd}
shows the CMD for the hard X-ray selected AGN in comparison with
optically selected galaxies from the COMBO-17 in the redshift interval
$0.2<z<1.2$. For clarity radio loud AGN are not plotted. X-ray sources
in this figure are preferentially associated with relatively luminous
galaxies evenly spread between the blue and the red clouds. This
suggests a mix of early and late-type hosts.

X-ray/radio matched AGN that scatter around the MIR/1.4\,GHz relation
of star-forming galaxies are all found in the blue cloud of Fig.
\ref{cmd}. This is further evidence for a young stellar population in
these systems that is also likely to dominate the radio luminosity. On
the contrary, X-ray/radio matched AGN without MIR counterparts lie in
the red cloud, suggesting old stellar populations and radio emission
associated with accretion on the central BH. These same sources also
deviate from the MIR/1.4\,GHz relation of star-forming galaxies,
consistent with their position in the CMD.

We caution for two effects that may bias the position of AGN in the
colour-magnitude. The red colour of some AGN may partly be a result of
dust extinction, while scattered/direct light from the AGN itself may
shift the rest-frame colours of some sources toward the blue
cloud. Visual inspection of HST/ACS images suggests that for some
systems this is indeed the case. Nevertheless, for most X-ray sources
in the sample the optical colours are dominated by the host galaxy
with small contamination from the central engine. A thorough analysis
of these effects will be presented in a future publication (Rovilos et
al. in preparation).

\subsection{Radio detection and X-ray obscuration}\label{obscuration}

  \begin{figure}
  \centering
  \includegraphics[width=8.8cm]{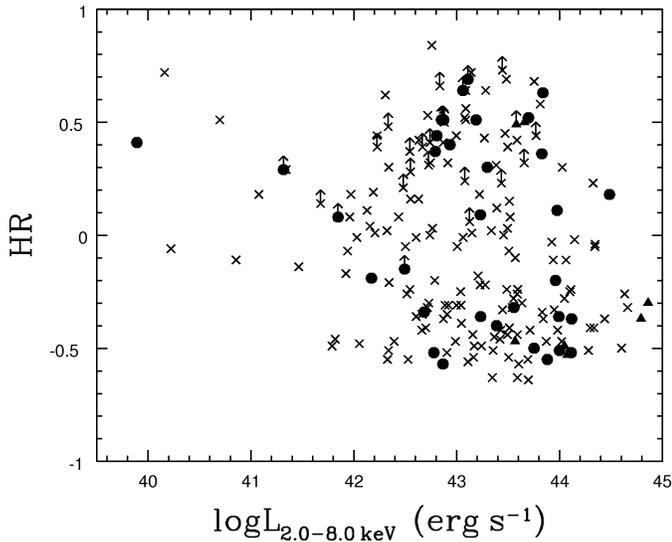}
  \caption{Hardness ratio against hard band X-ray luminosity. Filled
           circles are X-ray/radio matches from both the CDFS and
           the E-CDFS. Filled triangles are radio loud AGN
           $\left(\log L_{\rm 1.4\,GHz}>10^{25}\,\mathrm{W}\,\mathrm{Hz}^{-1}\right)$.
           Non-radio detected X-ray sources (CDFS only for clarity)
           are plotted with a cross.} 
  \label{hr-Lx}
  \end{figure}

In this section we explore a possible connection between radio
detection and X-ray obscuration suggested in previous studies
\citep[e.g.][]{Bauer2002,Georgakakis2004}. Figure \ref{hr-Lx} plots
the hardness ratio against the intrinsic hard X-ray luminosity. AGN
with radio counterparts are plotted with filled circles, while AGN not
detected in the radio are shown as crosses. For clarity, X-ray sources
without radio identifications are from the CDFS only. Filled triangles
mark radio loud sources.

\begin{figure}
  \centering
  \includegraphics[scale=0.25, clip=true]{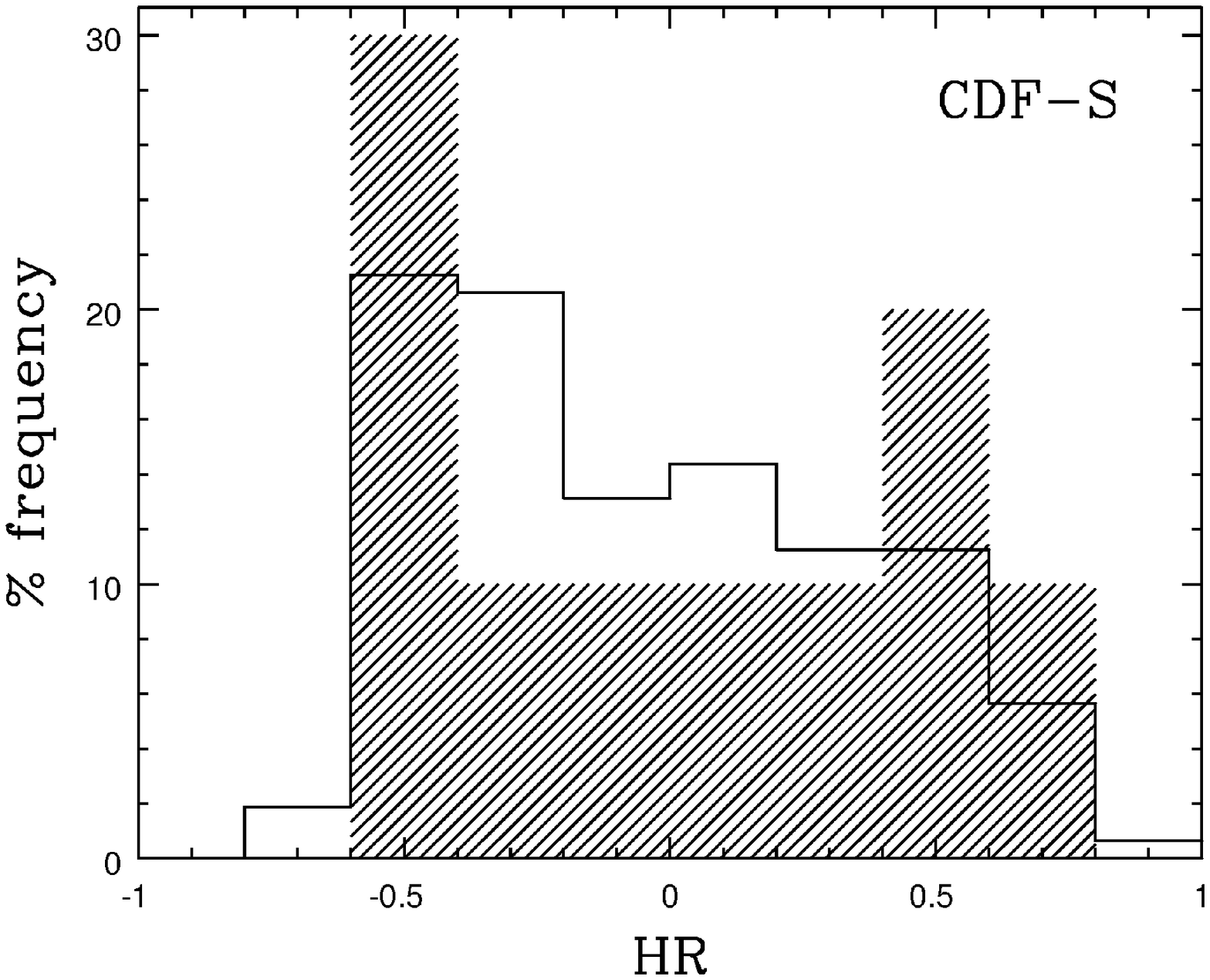}
  \includegraphics[scale=0.25, clip=true]{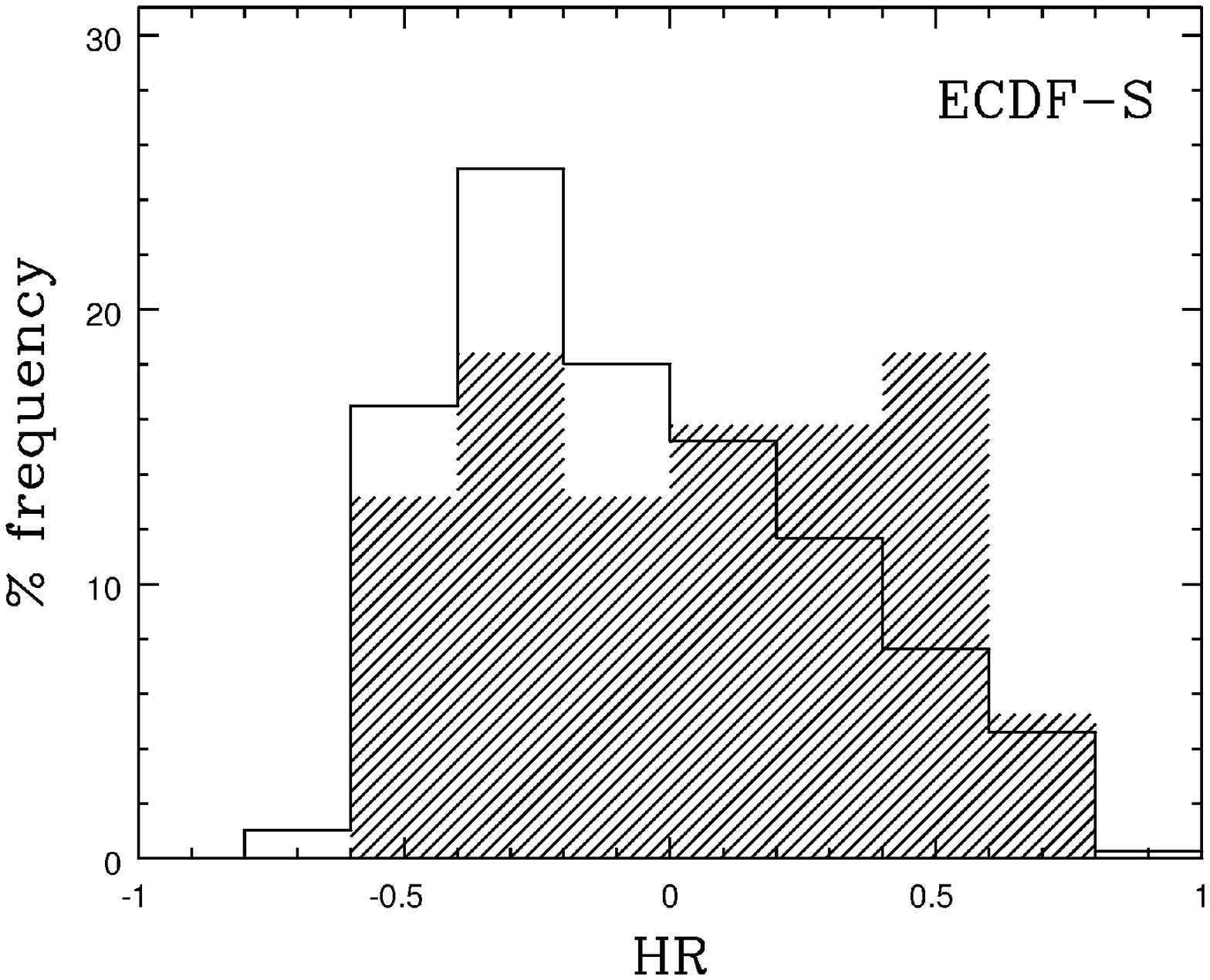}
  \caption{Hardness ratio distribution of X-ray sources with radio
           detections (shaded) and without radio counterparts. The left
           panel is for the CDFS and the right panel is for the E-CDFS.}
  \label{hist-hr}
  \end{figure}

\begin{figure}
  \centering
  \includegraphics[scale=0.24, clip=true]{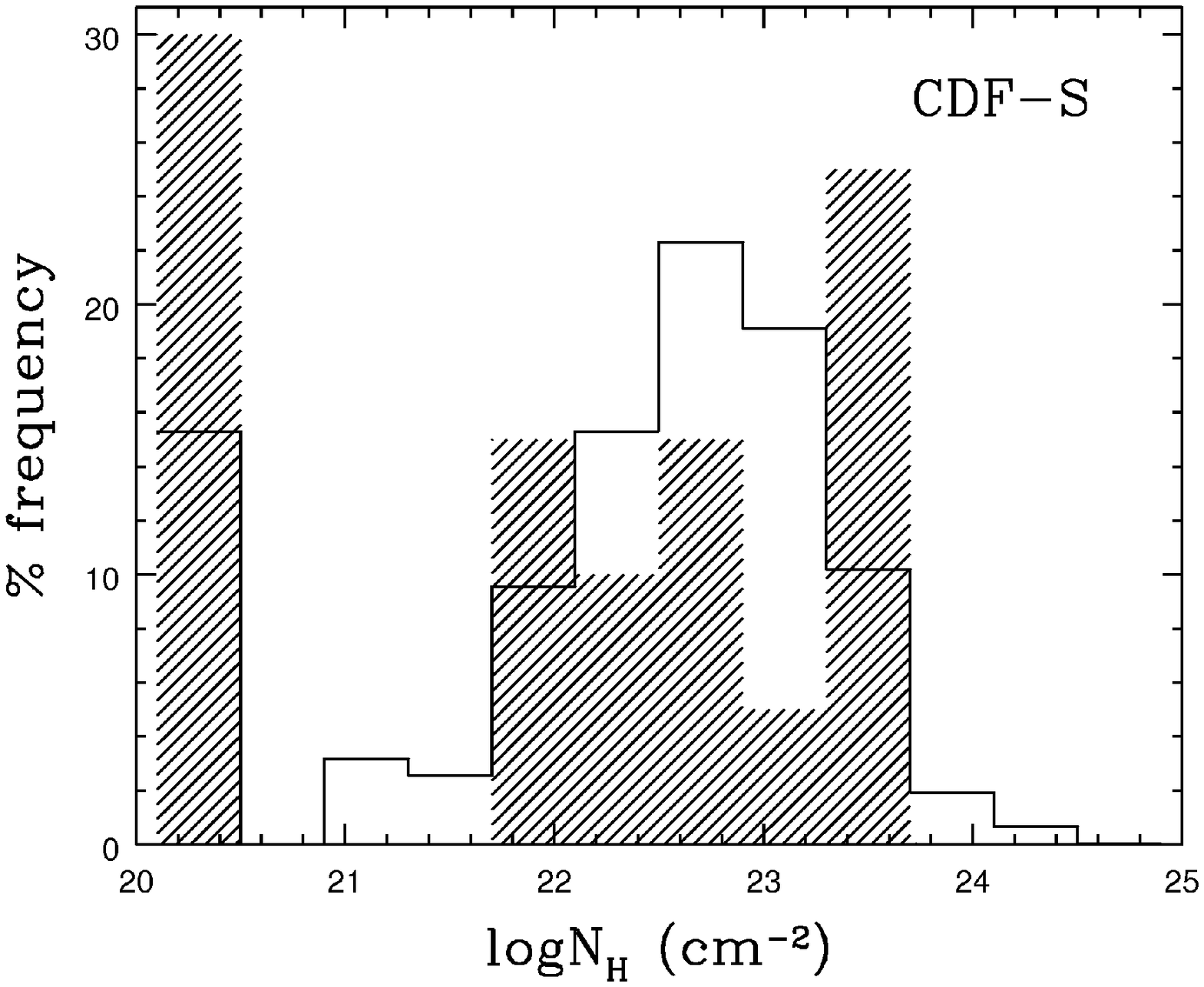}
  \includegraphics[scale=0.24, clip=true]{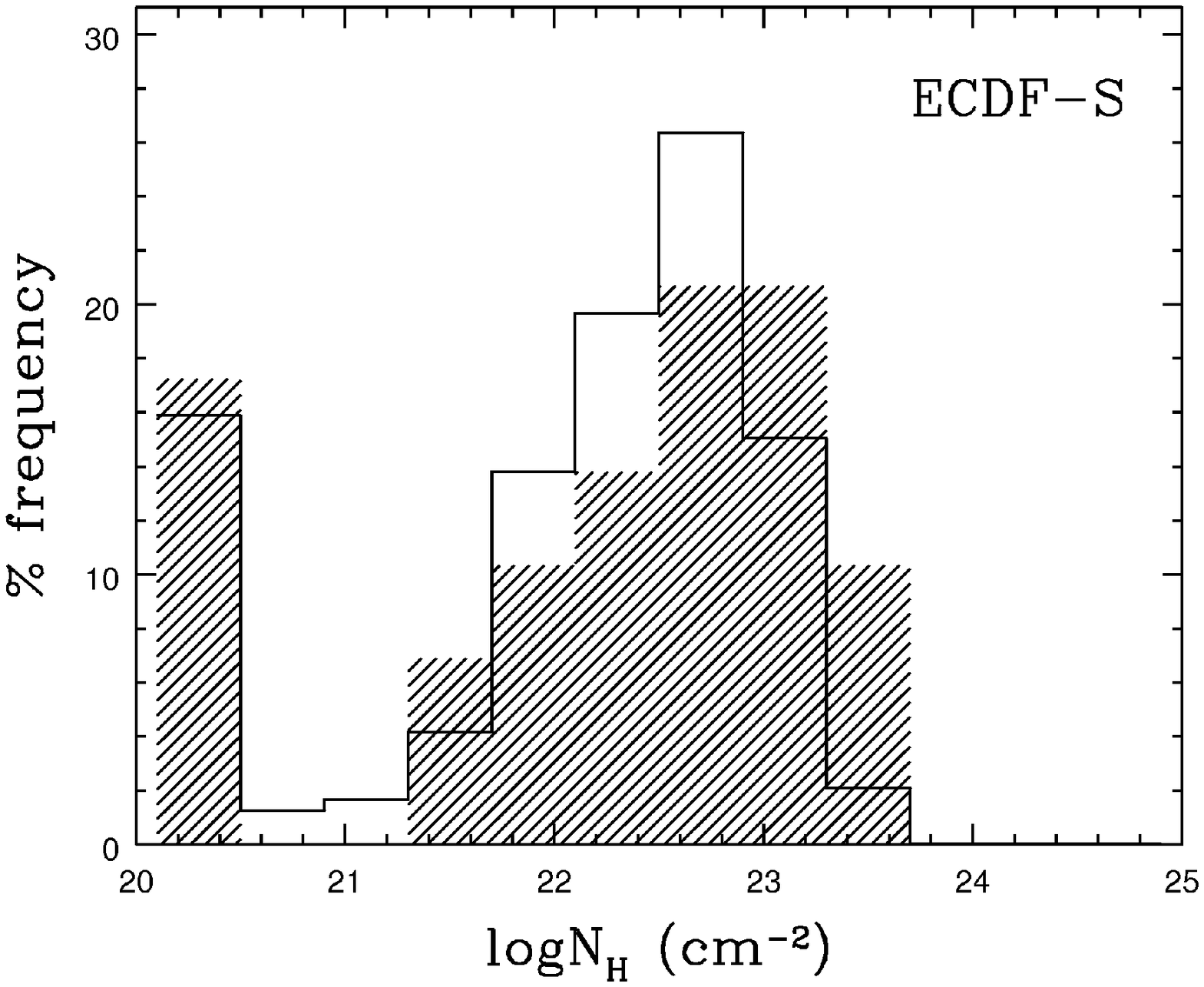}
  \caption{Hydrogen column density $N_{\mathrm{H}}$ histogram for radio
           detected (shaded) and not detected X-ray sources for the CDFS
           (left) and the E-CDFS (right).}
  \label{hist-nh}
  \end{figure}

In Fig. \ref{hr-Lx} there is no obvious tendency of radio detected
sources to be harder, in other words more obscured. This is further
demonstrated in Figs \ref{hist-hr} and \ref{hist-nh}, plotting the
$HR$ and $N_{\rm H}$ distributions respectively of X-ray sources with and
without radio identifications. We quantify the differences between
these two populations using the Gehan's statistical test as
implemented in the ASURV package \citep*{Isobe1986,Lavalley1992} to
take into account $HR$/$\rm N_H$ lower limits. This exercise does not
show any statistically significant differences in the obscuration
properties of X-ray sources with and without radio matches. The
probability that these two populations are drawn from the same parent
distribution is $>18\%$ for all the panels in Figs \ref{hist-hr}
and \ref{hist-nh}.

\section{Discussion}

\subsection{AGN - Starburst connection}

In the local Universe there is strong observational evidence, embodied
in the $M_{\mathrm{BH}}-\sigma$ relation
\citep[e.g.][]{Ferrarese2000,Gebhardt2000},
that the the build-up of the stellar mass of galaxies and the growth of
the super-massive BH at their centers are intimately related. At higher
redshift, $z\approx1$, close to the peak of the AGN density in the
Universe \citep[e.g.][]{Barger2005}, clear-cut examples of accreting BHs
associated with starburst events remain to be identified. The sensitivity
of \emph{Spitzer} in the mid-infrared has motivated studies that use
these wavelengths to explore the stellar population of AGN hosts
\citep[e.g.][]{Franceschini2005}. The main issue with any such
observational program however, relates to the difficulty in deconvolving
the AGN from the star formation emission. In this paper we follow a
multi-wavelength approach to explore the extent to which the faint
radio emission measures the level of star formation at $z\approx 1$

We have seen that a fraction (14\% and 11\% for the CDFS and E-CDFS
respectively) of the hard X-ray selected AGN have a radio
counterpart. Radio emission can be generated in a starburst
\citep[e.g.][]{Condon1991},
but it is also known to emerge from powerful quasars in the form of
radio jets and lobes \citep{Fanaroff1974}. Also low luminosity
AGN produce radio power as a result of their nuclear activity
\citep{Falcke2000,Nagar2002}.
We use the radio-infrared correlation to discriminate between the two
possible mechanisms generating radio emission, since there is a close
correlation between infrared and starburst
generated radio \citep*{Helou1985,Bressan2002}. In cases of a
starburst/AGN co-existence, \citet{Farrah2003} using SED fitting to
Ultra Luminous Infrared Galaxies, have shown that such a correlation is
attributed to the starburst part of the emission, while the infrared
is not correlated with the AGN generated radio. Therefore, the bulk of
the radio emission from a large fraction (up to 60\%) of our sample is
likely to be starburst powered. In addition to that, a large fraction of
the X-ray population not detected at 1.4\,GHz lies in the blue cloud
of the color-magnitude diagram and has radio luminosity upper limits
that are consistent with the $L_{\rm 24\mu m}-L_{\rm 1.4\,GHz}$ relation  for
starbursts. It is therefore possible that a large fraction of the
X-ray selected AGN at $z\approx1$ are associated with star-formation

The detection of star formation in X-ray selected AGN is an important result
and adds to the lines of evidence that there is a starburst-AGN relation.
These phenomena are often found in the same systems in cases of ultra-luminous
infrared galaxies \citep[see][]{Sanders1996} and also in other, optically
and X-ray selected galaxies \citep[e.g.][]{Kauffmann2003,Franceschini2005}.
While both phenomena have the
same dependency on large amounts of gas to function, there seems to be a more
fundamental link between them, and this gave rise to models establishing
the dependency of the star formation on the AGN properties through AGN feedback
\citep{Kauffmann2000,Hopkins2005,King2005,Croton2006}. Our results generally
support such models in the sense that a significant fraction of X-ray AGN
shows evidence of star formation. However, models requiring quenching
of the star-formation activity as a result of AGN feedback \citep{Hopkins2005}
predict that at the latest stages, AGN residing in the red cloud are
unobscured. This is not supported by our results, as there are no more
obscured sources in the blue cloud (not shown in Fig. \ref{cmd} for clarity
purposes). Alternatively, the AGN could be enhancing the starburst activity
instead of hindering it \citep{King2005}, or AGN cooling could be regulated
through a `radio mode' \citep{Croton2006} cutting the gas supply to the AGN.
This could also explain radio emission coming from the AGN and the positions
of such sources in the red cloud.

\subsection{X-ray obscuration}\label{xobs}

In the models mentioned in the previous paragraph, star formation plays
an important role in AGN evolution. Recent studies
\citep*[e.g.][]{Ballantyne2006} imply a connection between star formation
and obscuration in X-rays and we can test this connection using radio
emission as a probe of star formation activity.

Previous studies \citep{Bauer2002,Georgakakis2004} have
found a connection between those two properties, but we cannot confirm
those results. We first consider whether selection
effects could be responsible.
\citet{Bauer2002} use the 1\,Ms \emph{Chandra} Deep Field North
(CDF-N) survey which has detection limits very similar with the
CDFS but they use a deeper radio survey, reaching depths of
40\,$\mu$Jy in 1.4\,GHz, as opposed to
our limit of 60\,$\mu$Jy. As a result their radio
detection rate is higher and they report that a large fraction of
the X-ray - radio matches lie close to the detection limit and would
therefore be missed by our radio survey. They report that sources
classified as ``optically bright AGN'' (having
$-1<\log\left(\frac{f_{\mathrm{x}}}{f_{\mathrm{opt}}}\right)<1$
and optical magnitude $I<24$) appear harder if radio detected.
The statistical significance of this result is however poor,
as the Gehan's test points out that the two hardness ratio
distributions (for radio detected and non radio detected sources) are not
different with a  83\% probability. The small number of sources (13)
in this analysis however limits the statistical results one can derive.
In order to check if a larger sample would yield different results,
we used all hard selected X-ray sources from the 3\,arcmin sample
of \citet{Bauer2002} and performed the classification used in this
paper to select an AGN sample of 40 sources, 14 of which are radio identified.
Performing the same test we find that the null hypothesis is 70\% probable.

\citet{Georgakakis2004} on the other hand use similar radio limits
(60\,$\mu$Jy at 1.4\,GHz) with this study but have a shallower
X-ray survey, their sources have
$f_{\mathrm{x}}(2-8\,\mathrm{keV})>7.7\times10^{-15}\mathrm{erg}\,\mathrm{cm}^{-2}\,\mathrm{s}^{-1}$.
If we limit our sample to match the hard X-ray limits of
\citet{Georgakakis2004} we again find no correlation;
the Gehan's test gives a probability of 40\% that the radio detected
and non radio detected samples have from the same hydrogen column density
distribution. This number refers to the E-CDFS case,
where the sample is large enough (45 sources, 11 radio matches), so
that small number statistics do not affect our results in great extent. 
Although our numbers are similar to \citet{Georgakakis2004}
(43 sources, 14 radio matches), we cannot confirm their result.

The above result does not change if we limit our analysis to radio
sources that are consistent with the radio-infrared correlation.
Excluding sources with no infrared information from our sample and
repeating the statistical tests results in a null hypothesis probability
of 79\%. The fact that we do not find any correlation between X-ray
obscuration and radio emission means that the obscured
and non obscured sources are not different in terms of their radio
emission properties, and such an assumption is consistent with AGN unification
models \citep[e.g.][]{Antonucci1993}. Indeed if we consider the
standard scheme where absorption is attributed to a toroidal
structure round the AGN and assume that this torus is connected
with star formation \citep{Wada2002,CidFernandes2001}, then only weak
radio-obscuration correlation is expected because (starburst generated) radio
emission is considered isotropic from the torus, whereas X-ray
absorption is dependent on the line of sight. This setup explains
the unobscured radio sources seen in Fig \ref{hist-nh}.
Our results therefore support AGN orientation-based unification
models where the obscuring material is not symmetrically distributed,
in contrast to models which favor spherical covering of the central
engine \citep[e.g.][]{Fabian1998}.

  \begin{figure}
  \centering
  \includegraphics[width=8.8cm]{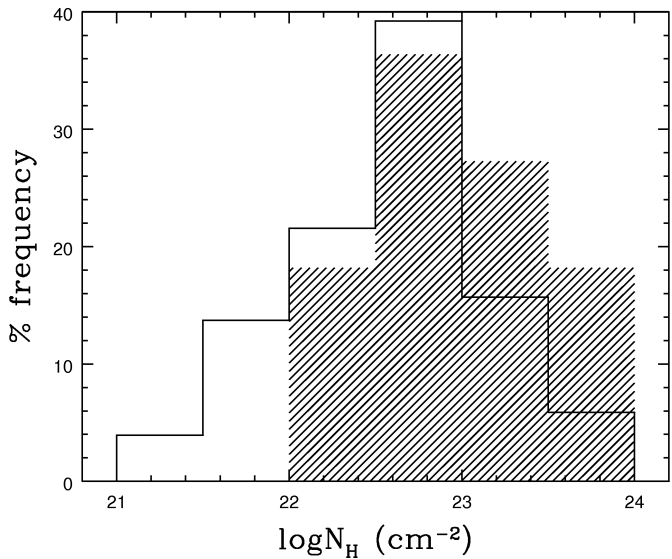}
  \caption{Hydrogen column density $(N_{\rm H})$ distributions for the
           subsample described in \S\ref{xobs}, i.e. hard X-ray
           selected radio quiet AGN which have an infrared counterpart
           and some degree of X-ray obscuration
           $(N_{\rm H}>10^{21}\,{\rm cm}^{-2})$. Shaded is the distribution for
           sources with a radio counterpart.}
  \label{final_hist}
  \end{figure}

If this the case, the lack of correlation reflects the existence
of sources hosting starburst activity, but the starburst clouds do not
lie in the line of sight, thus being unobscured in X-rays.
To check for this effect, we repeat the statistical test taking
into consideration only sources with some degree of obscuration
($N_{\mathrm{H}}>10^{21}\,\mathrm{cm}^{-2}$). This sub-sample
consists of 62 sources, 11 of which with a radio counterpart.
The $N_{\mathrm{H}}$ histograms for sources in this sub-sample with
and without a radio counterpart are shown in Fig. \ref{final_hist}
and the statistical results show that there is a 97\% probability that the
populations of radio detected and non radio detected
sources are \emph{different} in terms of $N_{\mathrm{H}}$. This supports the
assumption that both the AGN contribution to the radio emission and the
anisotropic distribution of the starburst (and radio emitting) clouds
are responsible for the lack of correlation between radio emission and
X-ray absorption.

\section{Conclusions}

We combine X-ray data of the southern \emph{Chandra} Deep fields
with radio observations of the same region. We aim to use the radio
as a proxy of star formation activity and explore its relation to
X-ray selected AGN. Optical and 24\,$\mu$m infrared observations
of the same region assist in defining the origin of the radio
emission, as they offer independent means of tracing star formation.
Our main results are summarized below:
\begin{itemize}
\item{The radio detection rate of X-ray sources is 14\%  and 9\% for the
      CDFS and the E-CDFS respectively, while 14\% and 14\% respectively
      of the hard selected sources have radio counterparts.}
\item{There is evidence that the radio emission of up to 60\% of the radio
      detected X-ray sources is linked to star formation.
      There is also a number of sources whose radio
      emission is attributed to the AGN; the radio-infrared correlation
      and the color distribution can identify them.}
\item{As derived by the $L_{\rm 24\,\mu m}-L_{\rm 1.4\,GHz}$ relation, we do not
      find strong evidence for the presence of numerous radio excess AGN
      which are not detected in the 1\,Ms CDFS exposure.}
\item{We found no strong evidence of correlation between radio emission
      and X-ray absorption. This is partly a result of the AGN being
      responsible for the radio emission of some sources and partly a result
      of the obscuring material being anisotropically distributed.}
\end{itemize}

\begin{acknowledgements}
ER acknowledges funding from the European Social Fund, Operational
Program for Educational and Vocational Training II (EPEAEK II), and
in particular the Program PYTHAGORAS II. AG acknowledges financial
support from the Marie-Curie Fellowship grant MEIF-CT-2005-025108.
JA gratefully acknowledges the support from the Science and Technology
Foundation (FCT, Portugal) through the research grant
POCI/CTE-AST/58027/2004.
\end{acknowledgements}

\begin{sidewaystable*}
\begin{minipage}[t][180mm]{\textwidth}
\caption{X-ray, radio and infrared properties of all X-ray
         sources matched with radio counterparts.}\label{table}
\centering
\begin{tabular}{cccccccccccc}
\hline\hline
A03 & L05 & $\alpha$ & $\delta$ & $z$ &         $f_{\mathrm{x}}\,(2-8)$\,keV             & $\log L_{\mathrm{x}}\,(2-8)$\,keV  & $HR$ & $\log N_{\mathrm{H}}$ & $\log L_{\rm 1.4\,GHz}$ & $\log L_{\rm 24\,\mu m}$ & Type \\
    &     &  (J2000) &  (J2000) &     & ($\times10^{-16}$\,erg\,cm$^{-2}$\,s$^{-1}$) &     (erg\,s$^{-1}$)     &      &       (cm$^{-2}$)    &   (W\,Hz$^{-1}$)        &      (W\,Hz$^{-1}$)    &      \\
\hline
12  & -   & 3:31:51.16 & -27:50:51.6 & 0.674       &     9.75 &    42.24 &    -0.29 & 21.96 & 23.09 & -     & AGN1        \\
27  & 240 & 3:31:57.71 & -27:42:08.4 & 0.654$^{1}$ &    28.94 &    42.68 &    -0.34 & 21.76 & 23.37 & -     & AGN1        \\
37  & 258 & 3:32:01.42 & -27:46:47.1 & 1.010$^{1}$ &    95.81 &    43.66 &     0.50 & 23.15 & 26.41 & 23.92 & AGN2        \\
48  & 280 & 3:32:04.89 & -27:41:27.6 & 0.720       &    33.94 &    42.85 &  $>$0.51 & 22.97 & 23.30 & 23.27 & AGN2        \\
57  & -   & 3:32:07.18 & -27:51:28.1 & 0.350$^{2}$ &  $<$3.41 & $<$41.11 & $<$-0.12 & 22.07 & 22.70 & -     & Galaxy      \\
65  & -   & 3:32:08.50 & -27:46:48.6 & 0.310       &     7.17 &    41.32 &  $>$0.29 & 22.48 & 22.76 & 23.46 & AGN2        \\
66  & 305 & 3:32:08.66 & -27:47:34.4 & 0.543       &   525.32 &    43.75 &    -0.50 & 20.00 & 24.32 & 24.58 & AGN1        \\
69  & 313 & 3:32:09.69 & -27:42:48.4 & 0.733       &  $<$4.82 & $<$42.02 & $<$-0.52 & 20.00 & 23.70 & -     & AGN1        \\
74  & -   & 3:32:10.80 & -27:46:27.6 & 1.013$^{1}$ &  $<$2.97 & $<$42.15 & $<$-0.05 & 22.61 & 24.34 & 23.91 & AGN1        \\
76  & 316 & 3:32:10.91 & -27:44:15.1 & 1.615       &    75.51 &    44.05 &    -0.49 & 20.00 & 25.63 & 24.87 & AGN1        \\
78  & 318 & 3:32:11.00 & -27:40:53.7 & 0.181       &    29.31 &    41.42 &    -0.40 & 21.07 & 22.43 & -     & Galaxy      \\
88  & 328 & 3:32:13.24 & -27:42:40.9 & 0.605       &   122.69 &    43.23 &     0.09 & 22.52 & 23.24 & 24.03 & AGN2        \\
109 & 345 & 3:32:16.21 & -27:39:30.4 & 1.324       &   141.21 &    44.11 &    -0.52 & 20.00 & 24.37 & 25.21 & AGN1        \\
117 & 348 & 3:32:17.14 & -27:43:03.3 & 0.569       &    50.11 &    42.78 &    -0.52 & 20.00 & 22.98 & 24.28 & AGN1        \\
118 & 347 & 3:32:17.18 & -27:52:20.9 & 1.097       &   117.82 &    43.84 &     0.63 & 23.31 & 23.56 & 24.36 & AGN2        \\
121 & -   & 3:32:18.07 & -27:47:18.2 & 0.734       &  $<$2.62 & $<$41.76 & $<$-0.58 & 20.00 & 23.95 & -     & Galaxy      \\
129 & -   & 3:32:19.81 & -27:41:23.1 & 0.229       &  $<$6.50 & $<$40.98 & $<$-0.15 & 21.91 & 22.53 & 23.67 & Galaxy      \\
144 & -   & 3:32:22.51 & -27:48:04.8 & 0.167$^{1}$ &  $<$2.22 & $<$40.22 & $<$-0.44 & 20.74 & 22.05 & -     & Galaxy      \\
148 & -   & 3:32:22.63 & -27:44:26.0 & 0.737       &  $<$3.47 & $<$41.89 &     0.12 & 22.64 & 23.33 & 24.15 & AGN2        \\
153 & -   & 3:32:23.88 & -27:58:42.4 & 0.123$^{1}$ & $<$16.94 & $<$40.81 &    -0.01 & 22.00 & 21.84 & -     & AGN2        \\
164 & -   & 3:32:25.17 & -27:54:49.6 & 1.090       &    22.45 &    43.11 &  $>$0.69 & 23.35 & 23.64 & 23.94 & AGN2        \\
177 & 379 & 3:32:27.00 & -27:41:05.1 & 0.734       &   539.43 &    44.07 &    -0.53 & 20.00 & 25.64 & 24.38 & AGN1        \\
181 & -   & 3:32:28.73 & -27:46:20.2 & 0.738       &     1.69 &    41.58 &    -0.67 & 20.00 & 23.72 & -     & AGN1$^{3}$  \\
189 & 390 & 3:32:29.88 & -27:44:25.0 & 0.076       &  $<$3.15 & $<$39.62 & $<$-0.79 & 20.00 & 22.18 & 23.01 & Galaxy      \\
193 & -   & 3:32:30.06 & -27:45:23.5 & 0.960       &    17.46 &    42.87 &    -0.57 & 20.00 & 23.68 & 24.34 & AGN1        \\
192 & 392 & 3:32:30.01 & -27:44:04.0 & 0.076       &     5.65 &    39.88 &    -0.56 & 20.00 & 21.80 & 22.80 & Galaxy      \\
197 & -   & 3:32:31.47 & -27:46:23.0 & 2.223       &  $<$2.57 & $<$42.91 &  $<$0.02 & 23.24 & 24.27 & 25.27 & AGN1        \\
219 & -   & 3:32:35.72 & -27:49:16.0 & 2.578       &     4.40 &    43.30 &     0.30 & 23.64 & 24.48 & 25.67 & AGN2        \\
228 & -   & 3:32:37.20 & -27:57:47.5 & 0.132$^{1}$ &  $<$9.95 & $<$40.65 &  $<$0.24 & 22.26 & 21.77 & -     & Galaxy      \\
230 & 414 & 3:32:37.77 & -27:52:12.4 & 1.603       &    68.48 &    44.00 &    -0.51 & 20.00 & 24.10 & 25.43 & AGN1        \\
240 & 420 & 3:32:38.92 & -27:57:00.4 & 0.304$^{1}$ &   230.43 &    42.81 &     0.44 & 22.59 & 22.49 & 23.99 & AGN2        \\
245 & 425 & 3:32:39.68 & -27:48:50.7 & 3.064       &    45.34 &    44.48 &     0.18 & 23.67 & 24.61 & 25.31 & AGN2        \\
247 & 433 & 3:32:40.84 & -27:55:46.6 & 0.625       &    57.44 &    42.93 &     0.40 & 22.82 & 23.18 & 24.05 & AGN2        \\
260 & 445 & 3:32:44.28 & -27:51:41.0 & 0.279       &  $<$2.86 & $<$40.82 & $<$-0.71 & 20.00 & 23.07 & 24.20 & Galaxy      \\
265 & -   & 3:32:44.98 & -27:54:38.7 & 0.458       &  $<$4.57 & $<$41.52 & $<$-0.09 & 22.19 & 22.94 & 23.82 & Galaxy      \\
270 & 460 & 3:32:46.76 & -27:42:12.6 & 0.103       &  $<$7.04 & $<$40.27 & $<$-0.63 & 20.00 & 21.53 & 22.17 & Galaxy      \\
274 & 464 & 3:32:47.86 & -27:42:32.8 & 0.979       &   206.58 &    43.96 &    -0.20 & 22.36 & 23.79 & 24.78 & AGN1        \\
278 & 469 & 3:32:49.23 & -27:40:49.8 & 1.222       &    50.15 &    43.58 &  $>$0.49 & 23.25 & 25.60 & 25.44 & AGN2        \\
\hline
\end{tabular}
\vfill
\end{minipage}
\end{sidewaystable*}

\setcounter{table}{0}
\begin{sidewaystable*}
\begin{minipage}[t][180mm]{\textwidth}
\caption{continued}
\centering
\begin{tabular}{cccccccccccc}
\hline\hline
A03 & L05 & $\alpha$ & $\delta$ & $z$ &         $f_{\mathrm{x}}\,(2-8)$\,keV             & $\log L_{\mathrm{x}}\,(2-8)$\,keV  & $HR$ & $\log N_{\mathrm{H}}$ & $\log L_{\rm 1.4\,GHz}$ & $\log L_{\rm 24\,\mu m}$ & Type \\
    &     &  (J2000) &  (J2000) &     & ($\times10^{-16}$\,erg\,cm$^{-2}$\,s$^{-1}$) &     (erg\,s$^{-1}$)     &      &       (cm$^{-2}$)    &   (W\,Hz$^{-1}$)        &      (W\,Hz$^{-1}$)    &      \\
\hline
284 & -   & 3:32:51.78 & -27:44:35.7 & 0.522       &  $<$4.40 & $<$41.64 & $<$-0.09 & 22.24 & 23.02 & 24.11 & Galaxy      \\
285 & 488 & 3:32:51.83 & -27:42:28.9 & 1.027       & $<$11.28 & $<$42.75 & $<$-0.11 & 22.54 & 23.90 & 24.61 & AGN1        \\
302 & 518 & 3:32:59.31 & -27:48:58.5 & 1.280$^{2}$ &    59.02 &    43.70 &     0.52 & 23.31 & 23.74 & 25.03 & AGN2        \\
315 & 544 & 3:33:02.97 & -27:51:46.4 & 3.690$^{2}$ &     7.32 &    43.88 &    -0.55 & 20.00 & 24.87 & 25.63 & AGN1        \\
325 & 592 & 3:33:09.48 & -27:46:03.4 & 0.347       &    39.06 &    42.17 &    -0.19 & 21.94 & 22.65 & -     & AGN2        \\
326 & 597 & 3:33:10.18 & -27:48:41.8 & 0.812$^{1}$ &    89.84 &    43.40 &    -0.40 & 21.54 & 25.78 & -     & AGN1        \\
-   & 7   & 3:31:15.04 & -27:55:18.6 & 0.494$^{1}$ &   284.67 &    43.39 &    -0.40 & 21.33 & 24.15 & -     & AGN1        \\
-   & 46  & 3:31:24.90 & -27:52:07.9 & 1.262$^{1}$ &    45.16 &    43.57 &    -0.47 & 20.00 & 26.51 & -     & AGN1        \\
-   & 47  & 3:31:25.29 & -27:59:58.7 & 0.789$^{1}$ &   255.87 &    43.83 &     0.36 & 22.90 & 23.73 & -     & AGN2        \\
-   & 51  & 3:31:27.23 & -27:42:46.9 & -           &    15.59 &    -     &    -0.18 & -     & -     & -     & -           \\
-   & 66  & 3:31:30.07 & -27:56:02.3 & 0.690$^{1}$ &  $<$7.34 & $<$42.14 & $<$-0.27 & 22.04 & 24.34 & -     & AGN1        \\
-   & 74  & 3:31:31.63 & -27:45:19.2 & 0.138$^{1}$ &  $<$9.99 & $<$40.69 & $<$-0.13 & 21.85 & 20.91 & -     & Galaxy      \\
-   & 82  & 3:31:32.82 & -28:01:16.1 & 0.148$^{1}$ &  $<$9.77 & $<$40.75 & $<$-0.13 & 21.86 & 22.04 & -     & Galaxy      \\
-   & 94  & 3:31:35.43 & -28:03:15.7 & 0.083$^{1}$ &    42.74 &    40.84 &    -0.64 & 20.00 & 21.48 & -     & Galaxy      \\
-   & 136 & 3:31:43.21 & -27:54:05.3 & -           &    95.28 &    -     &     0.34 & -     & -     & -     & -           \\
-   & 141 & 3:31:43.48 & -27:51:03.0 & 0.265$^{1}$ & $<$11.79 & $<$41.38 &  $<$0.10 & 22.25 & 22.26 & -     & Galaxy      \\
-   & 146 & 3:31:44.04 & -27:38:35.7 & 0.057$^{1}$ &    10.37 &    39.89 &     0.41 & 22.33 & 21.29 & -     & AGN2        \\
-   & 177 & 3:31:48.57 & -28:04:32.9 & -           &  $<$9.05 &    -     &    -0.10 & -     & -     & -     & -           \\
-   & 205 & 3:31:52.16 & -27:39:26.0 & -           &  $<$7.59 &    -     &  $<$0.05 & -     & -     & -     & -           \\
-   & 254 & 3:32:00.83 & -27:35:56.6 & 0.948$^{1}$ &    41.64 &    43.23 &    -0.36 & 21.89 & 24.90 & -     & AGN1        \\
-   & 268 & 3:32:03.04 & -27:35:16.5 & 0.000       &     0.00 &    0.03  &    -0.03 & -     & -     & -     & Star$^{4}$  \\
-   & 282 & 3:32:05.21 & -28:04:14.7 & -           &    69.15 &    -     &     0.02 & -     & -     & -     & -           \\
-   & 289 & 3:32:06.09 & -27:32:36.1 & -           &   142.70 &    -     &    -0.22 & -     & -     & -     & -           \\
-   & 321 & 3:32:11.64 & -27:37:26.0 & 1.574$^{1}$ &   521.28 &    44.86 &    -0.30 & 22.41 & 25.75 & -     & AGN1        \\
-   & 385 & 3:32:28.56 & -27:35:37.0 & 0.677$^{1}$ &   717.60 &    44.12 &    -0.37 & 21.65 & 23.25 & -     & AGN1        \\
-   & 398 & 3:32:32.01 & -28:03:09.7 & 1.966$^{1}$ &   261.23 &    44.79 &    -0.37 & 22.29 & 26.79 & -     & AGN1        \\
-   & 504 & 3:32:56.49 & -27:58:48.2 & 0.154$^{1}$ &     5.74 &    40.55 &    -0.33 & 21.41 & 22.93 & -     & Galaxy      \\
-   & 506 & 3:32:57.00 & -27:33:43.7 & -           &    50.59 &    -     &     0.33 & -     & -     & -     & -           \\
-   & 520 & 3:32:59.54 & -28:01:23.7 & 2.386$^{1}$ &    26.19 &    43.99 &    -0.36 & 22.52 & 24.87 & -     & AGN1        \\
-   & 538 & 3:33:01.46 & -28:02:51.4 & 1.265$^{1}$ &   115.23 &    43.98 &     0.11 & 22.94 & 24.11 & -     & AGN2        \\
-   & 552 & 3:33:03.76 & -27:36:11.0 & 1.574$^{1}$ &    25.85 &    43.56 &    -0.32 & 22.37 & 24.69 & -     & AGN1        \\
-   & 555 & 3:33:04.43 & -27:38:01.5 & 0.991$^{1}$ &  $<$9.66 & $<$42.64 &    -0.09 & 22.55 & 23.68 & -     & AGN2        \\
-   & 557 & 3:33:05.13 & -27:40:27.6 & 0.302$^{1}$ &  $<$3.68 & $<$41.01 & $<$-0.19 & 21.90 & 22.46 & -     & Galaxy      \\
-   & 581 & 3:33:08.16 & -27:50:33.2 & 0.730$^{1}$ &    34.32 &    42.87 &  $>$0.51 & 22.98 & 24.43 & -     & AGN2        \\
-   & 587 & 3:33:09.10 & -27:58:46.2 & 0.674$^{1}$ &  $<$7.37 & $<$42.12 &  $<$0.21 & 22.68 & 23.04 & -     & AGN1        \\
-   & 599 & 3:33:10.28 & -27:33:06.5 & -           &    16.09 &    -     &  $>$0.24 & -     & -     & -     & -           \\
-   & 609 & 3:33:11.77 & -27:41:38.6 & 1.059$^{1}$ &     5.82 &    42.49 & $>$-0.15 & 22.50 & 24.17 & -     & AGN2        \\
-   & 632 & 3:33:16.14 & -28:02:21.0 & 0.362$^{1}$ &    16.76 &    41.85 &  $>$0.08 & 22.32 & 22.89 & -     & AGN2        \\
\hline
\end{tabular}
\vfill
\end{minipage}
\end{sidewaystable*}

\setcounter{table}{0}
\begin{sidewaystable*}
\begin{minipage}[t][180mm]{\textwidth}
\caption{continued}
\centering
\begin{tabular}{cccccccccccc}
\hline\hline
A03 & L05 & $\alpha$ & $\delta$ & $z$ &         $f_{\mathrm{x}}\,(2-8)$\,keV             & $\log L_{\mathrm{x}}\,(2-8)$\,keV  & $HR$ & $\log N_{\mathrm{H}}$ & $\log L_{\rm 1.4\,GHz}$ & $\log L_{\rm 24\,\mu m}$ & Type \\
    &     &  (J2000) &  (J2000) &     & ($\times10^{-16}$\,erg\,cm$^{-2}$\,s$^{-1}$) &     (erg\,s$^{-1}$)     &      &       (cm$^{-2}$)    &   (W\,Hz$^{-1}$)        &      (W\,Hz$^{-1}$)    &      \\
\hline
-   & 639 & 3:33:16.91 & -27:41:21.8 & 0.143$^{1}$ & $<$17.99 & $<$40.98 & $<$-0.15 & 21.83 & 22.16 & -     & Galaxy      \\
-   & 646 & 3:33:17.74 & -27:59:06.2 & 1.128$^{1}$ & $<$11.24 & $<$42.85 &  $<$0.25 & 23.00 & 24.00 & -     & AGN1        \\
-   & 657 & 3:33:19.09 & -27:35:30.9 & 0.139$^{1}$ &  $<$9.04 & $<$40.65 & $<$-0.24 & 21.65 & 22.00 & -     & Galaxy      \\
-   & 664 & 3:33:20.55 & -27:49:10.9 & 0.139$^{1}$ & $<$13.86 & $<$40.83 & $<$-0.69 & 20.00 & 21.93 & -     & Galaxy      \\
-   & 674 & 3:33:21.29 & -27:41:38.6 & 1.151$^{1}$ & $<$13.69 & $<$42.95 & $<$-0.03 & 22.71 & 24.40 & -     & AGN1        \\
-   & 676 & 3:33:22.75 & -27:54:59.1 & 0.831$^{1}$ &    20.68 &    42.79 &     0.37 & 22.92 & 23.38 & -     & AGN2        \\
-   & 738 & 3:33:34.56 & -27:47:51.0 & 0.635$^{1}$ &    99.40 &    43.19 &     0.51 & 22.91 & 23.93 & -     & AGN2        \\
-   & 743 & 3:33:36.39 & -27:43:55.1 & 1.114$^{1}$ & $<$17.75 & $<$43.03 &     0.06 & 22.80 & 24.30 & -     & AGN2        \\
-   & 746 & 3:33:36.65 & -27:42:23.6 & -           &    15.02 &    -     &  $>$0.21 & -     & -     & -     & -           \\
-   & 752 & 3:33:38.48 & -28:02:53.9 & -           & $<$20.86 &    -     & $<$-0.32 & -     & -     & -     & -           \\
-   & 118 & 3:31:39.99 & -27:41:57.2 & -           &   221.99 &    -     &    -0.02 & -     & -     & -     & -           \\
-   & 124 & 3:31:40.99 & -27:44:35.0 & -           &    60.28 &    -     &     0.06 & -     & -     & -     & -           \\
-   & 606 & 3:33:11.32 & -27:43:11.9 & 1.051$^{1}$ &    21.81 &    43.06 &  $>$0.64 & 23.29 & 23.76 & -     & AGN2        \\
\hline
\\
\multicolumn{12}{l}{\hspace{0.33cm}Notes:}\\[5pt]
\multicolumn{12}{l}{1. Photometric redshift from \citet{Wolf2004}}\\[2pt]
\multicolumn{12}{l}{2. Photometric redshift from \citet{Zheng2004}}\\[2pt]
\multicolumn{12}{l}{3. Classified as ``galaxy'' in \citet{Bauer2004}}\\[2pt]
\multicolumn{12}{l}{4. Galactic star, identified by optical photometry and spectroscopy \citep{Zheng2004,Szokoly2004}}
\end{tabular}
\vfill
\end{minipage}
\end{sidewaystable*}

\end{document}